\documentclass[aps,prb,twocolumn,superscriptaddress]{revtex4-2}
\usepackage[T1]{fontenc}
\usepackage{graphicx}
\usepackage{dcolumn}
\usepackage{bm}
\usepackage{amsmath}
\usepackage{amssymb}
\usepackage{array}
\usepackage{color}
\usepackage{xcolor}
\newcolumntype{P}[1]{>{\centering\arraybackslash}p{#1}}
\usepackage[colorlinks=true, breaklinks=true, linkcolor=blue, citecolor=blue, urlcolor=blue]{hyperref}
\newcommand{\CsCoBr}{Cs$_2$CoBr$_4$\xspace}
\newcommand{\AMX}{A$_2$MX$_4$\xspace}
\newcommand{\CsCuCl}{Cs$_2$CuCl$_4$\xspace}
\newcommand{\CsCuBr}{Cs$_2$CuBr$_4$\xspace}
\newcommand{\CsCoI}{Cs$_2$CoI$_4$\xspace}
\newcommand{\mb}[1]{\mathbf{#1}}
\newcommand{\mo}{\mu_0}

\newcommand{\ethz}{Laboratory for Solid State Physics, ETH Z\"{u}rich, 8093 Z\"{u}rich, Switzerland}
\newcommand{\dresden}{Dresden High Magnetic Field Laboratory (HLD-EMFL) and Würzburg-Dresden Cluster of Excellence ctd.qmat,
	Helmholtz-Zentrum Dresden-Rossendorf (HZDR), 01328 Dresden, Germany}

\usepackage{ifthen}
\usepackage{booktabs}
\usepackage{multirow}

\newcommand{\aver}[1]{\left\langle #1 \right\rangle}

\newcommand{\ket}[1]{\left| #1 \right\rangle}

\newcommand{\dd}{\mathrm{d}}

\newcommand{\mub}{\mu_\mathrm{B}}

\usepackage{gensymb,bbold,appendix,xspace}

\begin{document}

	\title{Counting kinks and magnons with the lattice: magnetoelastic effects in \CsCoBr
		}
	
	\date{\today}
	
	\begin{abstract}
	We report low-temperature dilatometry and ultrasound measurements on the frustrated quasi-one-dimensional antiferromagnet \CsCoBr across the entire cascade of field-induced magnetic phase transitions. The whole phase diagram can be recovered from these lattice probes, with distinct anomalies observed at each phase boundary. Elongation along the principal chain axis is found to closely track magnetization throughout the phase diagram. This behavior is interpreted in terms of kinks in the N\'eel structure and magnons in the transverse-ordered magnetic-BEC phases, each carrying a fixed lattice distortion. A simultaneous analysis of ultrasound and striction data allows us to quantify the strength of magnetoelastic coupling.
	\end{abstract}
	
	\author{V. Romerio}
	\affiliation{\ethz}
	\author{J. Nagl}
	\affiliation{\ethz}
	\author{K. Yu. Povarov}
	\affiliation{\dresden}
	\author{S. Gvasaliya}
	\affiliation{\ethz}
	\author{A. Zheludev}
	\affiliation{\ethz}

\maketitle
\section{Introduction}

Magnetoelastic coupling often plays an important role in frustrated quantum magnetic materials. 
In some cases the appearance of lattice distortions partially relieves degeneracies caused by frustration \cite{TchernyshyovChern2011_FrustratedMagnetism_SpringerChapter}, stabilizing particular ordered phases and sometimes even driving transitions between them \cite{Tsurkan2017_SciAdv3_e1601982_MnCr2S4_plateau_supersolid}. Among the more famous examples are pyrochlore \cite{Tchernyshyov2002_PRL88_067203,Penc2004_PRL93_197203,Matsuda2007_NatPhys3_397} and triangular-lattice compounds \cite{WangVishwanath2008_PRL100_077201}, as well as the Shastry--Sutherland system SrCu$_2$(BO$_3$)$_2$~\cite{Zherlitsyn2000_PRB62_R6097_SrCu2BO32_ultrasound,Jaime2012_PNAS109_12404_SrCu2BO32_100T_final,Radtke2015_PNAS112_1971_SrCu2BO32_nanopantograph}. Even for phases and phase transitions of purely magnetic origin, magnetoelastic coupling can lead to exciting new physics. In particular, strain~\cite{Gass2020_PRB101_245158_RuCl3}, optical-phonon~\cite{Sushkov2005_PRL94_137202}, and sound-propagation~\cite{Zherlitsyn2014_LTP40_123_spinlattice_AFM_review} anomalies are a common occurrence at magnetic phase transitions \cite{Luthi2007}. In certain cases they have even been used to directly measure two-spin correlation functions, as, for example, in the $S=1$ quantum paramagnet NiCl$_2\cdot$4SC(NH$_2$)$_2$ (DTN)~\cite{Zapf2008_PRB77_020404_authorcopy}, the spin ladder (C$_5$H$_{12}$N)$_2$CuBr$_4$ (BPCB)~\cite{Anfuso2008_PRB77_235113}, or SrCu$_2$(BO$_3$)$_2$~\cite{Bettler2020_PRR2_012010}.
Magnetoelastic coupling caused by exchange striction is particularly relevant in materials undergoing a cascade of field-induced phase transitions, especially those involving collinear plateau phases, the ``lattice staircase'' in CuFeO$_2$ being a particularly well-known example \cite{Terada2006_PRB74_180404_CuFeO2_lattice_staircase}.

\begin{figure}[h]
	\centering
	\includegraphics[width = \linewidth]{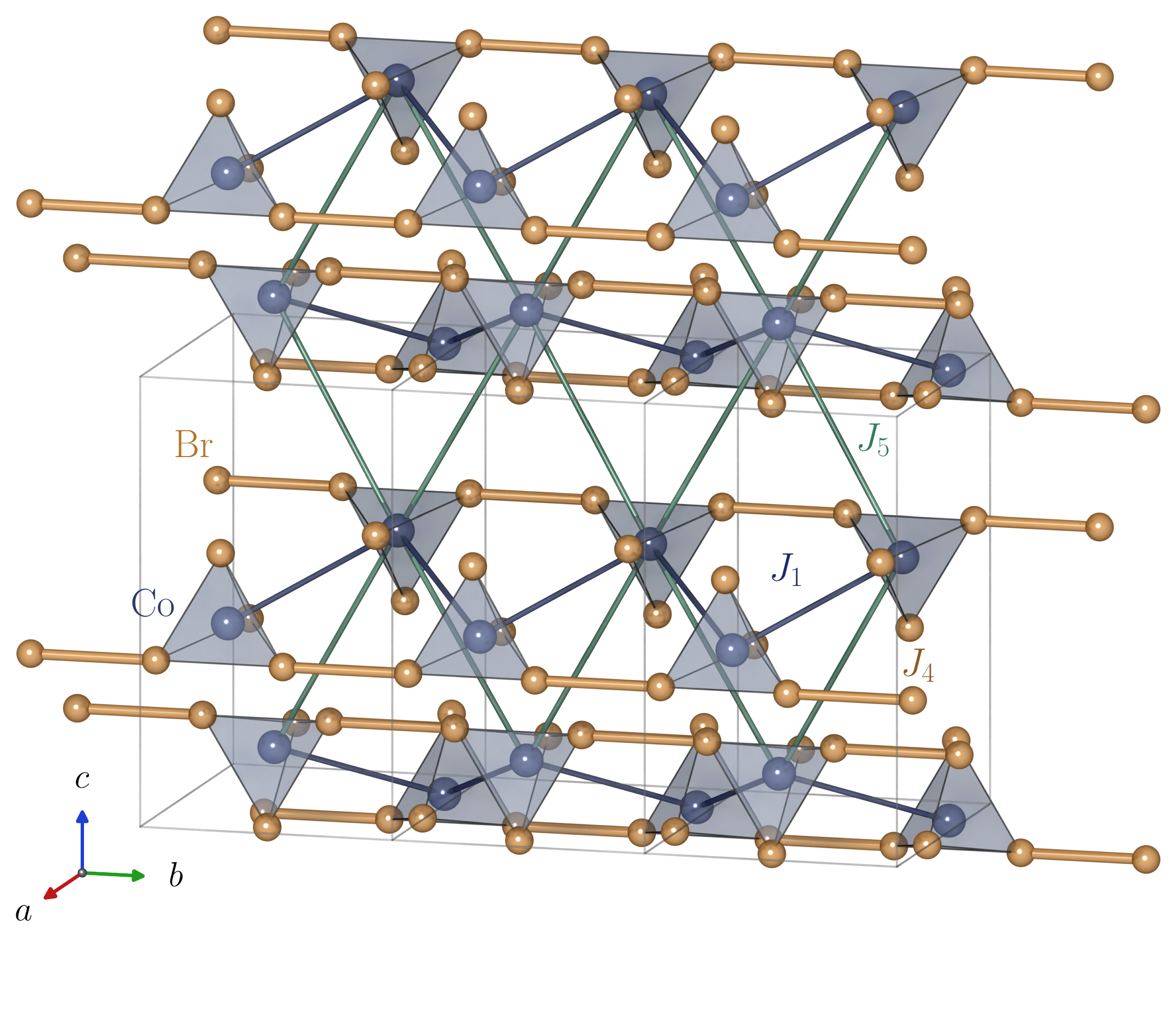}
	\caption{Dominant exchange interactions in \CsCoBr. The strongest coupling $J_4$ runs over a Co--Br--Br--Co superexchange path that involves the shortest Br--Br bond (golden) between CoBr$_4$ coordination tetrahedra (light blue). It forms Co$^{2+}$ chains running along the $b$ axis. These are pairwise bound into zigzag spin ladders by $J_1$. Four such ladders stacked along the $c$ axis are shown. The much weaker interaction $J_5$ diagonally couples nearby chains in the $(b,c)$ plane.}
	\label{Fig:Material_Structure}
\end{figure}

Cascades of field-induced plateau states and other phases are known in several frustrated quantum magnets. Particular diversity is found among orthorhombic (space group P$nma$) \AMX materials, where M is a magnetic transition metal, A an alkali metal, and X either a halogen or oxygen \cite{Cs2RuO4_NabiPRB2025}. The best-known species with complex field-induced phases are \CsCuCl \cite{Tokiwa2006_PRB73_134414_Cs2CuCl4_phase_diagram} and \CsCuBr \cite{Ono2003_PRB67_104431_Cs2CuBr4_plateau,Fortune2009_PRL102_257201_Cs2CuBr4_cascade}. The extent to which these transitions involve the crystal lattice remains poorly understood, but is expected to be significant. Indeed, high-pressure experiments on \CsCuCl show that magnetic interactions are sensitive to pressure \cite{Zvyagin2019_NatCommun10_1064}. A direct observation of magnetoelastic effects has recently been reported for \CsCoI \cite{Cs2CoI4_NabiPRM2026}. The focus of the present work is \CsCoBr. It stands out in that it exhibits one of the most complex cascades of field-induced phases~\cite{Cs2CoBr4_PovarovPRR2020} and is simultaneously one of the best characterized and well-understood species. Moreover, there is initial experimental evidence of strong magnetoelastic coupling in this material \cite{Cs2CoBr4_FacherisPRB2024}.

Over time, our understanding of \CsCoBr has been evolving \cite{Cs2CoBr4_PovarovPRR2020,Cs2CoBr4_FacherisPRL2022,Cs2CoBr4_FacherisPRL2023}, but by now its magnetic Hamiltonian has been solidly established through neutron scattering and THz spectroscopy experiments \cite{Cs2CoBr4_FacherisPRB2024} and numerical modeling \cite{Cs2CoBr4_HalatiPRR2026}. Magnetism is due to $S=3/2$ Co$^{2+}$ ions subject to dominant XYZ anisotropy with the crystallographic $b$ direction being the ``easiest'' axis. This anisotropy is of predominantly easy-plane character with an anisotropy constant $D = 0.70$~meV \cite{Cs2CoBr4_FacherisPRB2024}. As a consequence, the single-ion ground state responsible for most of the low-energy physics is an effective pseudospin $S=1/2$ doublet \cite{Cs2CoBr4_PovarovPRR2020}. The easy-axis role of the $b$ axis is further enhanced by the fact that the structure contains two crystallographically equivalent sets of magnetic ions with almost orthogonal anisotropy planes that share this direction. Compared to single-ion anisotropy, inter-ion interactions are considerably weaker. The dominant Heisenberg exchange coupling is antiferromagnetic, with $J_4 = 0.135(5)$~meV. It involves the fourth-shortest Co--Co distance and builds antiferromagnetic chains that also run along $b$. Next in the hierarchy is the nearest-neighbor Heisenberg coupling $J_1 = 0.06(1)$~meV that binds pairs of adjacent chains with parallel anisotropy planes into frustrated zigzag ladders. This geometry is illustrated in Fig.~\ref{Fig:Material_Structure}. The most spectacular manifestation of its predominantly one-dimensional character is the zero-field excitation spectrum, a Zeeman ladder of spinon bound states observed by neutron spectroscopy and studied in detail numerically \cite{Cs2CoBr4_FacherisPRL2023,Cs2CoBr4_HalatiPRR2026}. Indeed, other magnetic interactions have also been identified but are weaker still. The strongest among them is $J_5 = 0.020(5)$~meV, which diagonally connects adjacent chains in the $(b,c)$ plane. A more detailed discussion can be found in Ref.~\cite{Cs2CoBr4_FacherisPRB2024}.

From the above it is clear that the crystallographic $b$ axis is in many ways special for the magnetism of this compound. Applying a magnetic field along that direction indeed results in a complex cascade of five distinct ordered phases \cite{Cs2CoBr4_PovarovPRR2020,Cs2CoBr4_FacherisPRB2024}: a collinear N\'eel state ($\mo H = 0$--$1.5$~T), an incommensurate longitudinal spin density wave (SDW) with the propagation vector scaling proportionately to the pseudospin magnetization ($\mo H = 1.5$--$2.7$~T) \cite{Cs2CoBr4_FacherisPRL2022}, a collinear ``UUD'' pseudospin-plateau state ($\mo H = 2.7$--$4.0$~T), a spin-flop N\'eel phase ($\mo H = 4.0$--$4.5$~T) and finally an incommensurate fan phase with a field-independent propagation vector ($\mo H = 4.5$--$5.2$~T). Beyond that is a collinear pseudospin-saturated state. All but the final saturation transition are discontinuous. This again hints at strong magnetoelastic coupling that is known to drive and reinforce first-order magnetic transitions \cite{BeanRodbell1962_PR126_104,LarkinPikin1969_JETP29_891}. What makes \CsCoBr a unique platform for a {\em quantitative} study of these phenomena across all five phases is that their magnetic structures have been determined through single-crystal neutron diffraction experiments \cite{Cs2CoBr4_FacherisPRB2024,Cs2CoBr4_FacherisPRL2022}. In the present work we report dilatometry and ultrasound measurements through the entire phase diagram. We further demonstrate that the observed behavior along the chain-axis can be quantitatively explained through exchange striction of a single bond, namely $J_4$.

\begin{figure}[]
	\centering
	\includegraphics[width = 1\linewidth]{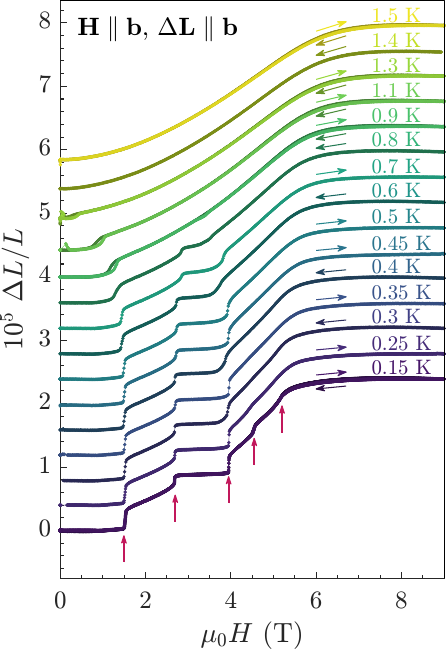}
	\caption{Relative isothermal elongation along the crystallographic $b$-axis against magnetic field applied along the same direction. Curves corresponding to successive temperatures are shifted by $\Delta L/L = 0.4\times 10^{-5}$ for visibility. The arrows on the right indicate the field sweep direction. Red vertical arrows indicate the low-temperature phase boundaries.}
	 \label{Fig:Data_Hiib}
\end{figure}

\section{Methods}
All measurements were performed on single crystals of \CsCoBr grown using the Bridgman method \cite{Cs2CoBr4_PovarovPRR2020}. Dilatometry data were collected using a commercial K\"{u}chler capacitance dilatometer \cite{Kuechler2012_RSI83_095102_dilatometer,Kuechler2023_RSI94_045108}. The samples were thin plates cut from as-grown crystals. The sample environment was a Quantum Design $^3$He-$^4$He dilution refrigerator PPMS insert. 
Capacitance was measured using an Andeen-Hagerling 2700A bridge operated at 1111 Hz. The realistic resolution of this setup is in the ballpark of 0.1~\AA~\cite{Kuechler2017_RSI88_083903_dilatometer_PPMS}.  Elongation was separately measured along the $b$- and $c$-directions, using 1.52~mm and 1.06~mm thick samples, respectively. In all cases the magnetic field was applied along the crystallographic $b$-axis. The data were collected at fixed temperatures between 0.15 and 1.5~K while sweeping the field between 0 and 9~T. The field sweep rate was chosen between $2\times 10^{-4}$~T/s and $10^{-3}$~T/s, with slower sweeps at lower temperatures to avoid induction heating of the dilatometer body.

 Ultrasound velocity and attenuation were measured using the pulse-echo technique with phase-sensitive lock-in detection. This technique is sensitive to the relative change in sound velocity $\Delta v/v$ of at least $10^{-5}$, and to changes in attenuation $\Delta\alpha$ of 0.1~dB/cm. No correction for sample striction was applied, since the observed relative velocity changes were at least 20 times larger than the measured relative elongation.
 Detailed data for the longitudinal acoustic $c_{33}$ mode (propagation and polarization along the crystallographic $c$-axis) were measured in a $1.580(5)$~mm long polished single-crystal sample with PVDF-foil transducers bonded with Flex + Bond to its opposing faces. The measurements were performed using a Zurich Instruments UHFLI lock-in amplifier at an excitation frequency of $26$~MHz. The sample environment was the same as in the dilatometry experiments. More limited $c_{22}$-ultrasound measurements (longitudinal sound along the $b$-axis) were performed in a $^3$He-$^4$He dilution refrigerator with an $18$~T magnet. Here Y-cut LiNbO$_3$ transducers were used, attached with Thiokol-32 to the polished faces of a $1.95(1)$~mm long crystal. The data were collected using a Zurich Instruments GHFLI lock-in amplifier at $37.4$~MHz. The magnetic field was applied along the $b$-axis at all times and swept at a rate of $1$--$2\times 10^{-3}$~T/s in measurements performed between 0.1 and 1.5~K. At select phase boundaries, slow $10^{-4}$~T/s sweeps in both directions were used to check for hysteretic behavior.

\section{Results}
\label{sec:data}

\subsection{Dilatometry}

\begin{figure}[]
	\centering
	\includegraphics[width = 1\linewidth]{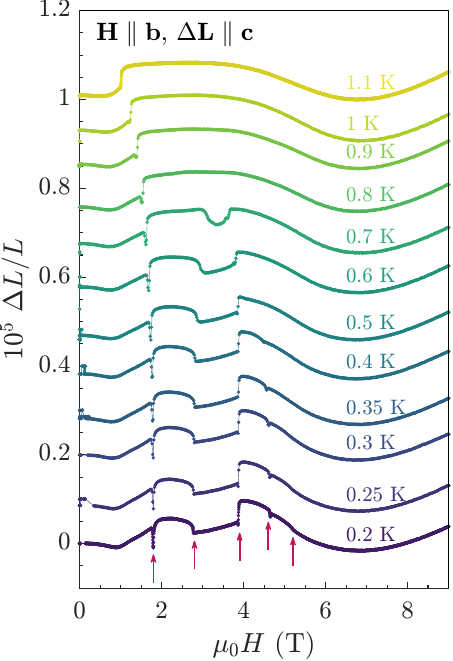}
	\caption{Same as Fig.~\ref{Fig:Data_Hiib}, for elongation along $c$ in a $b$-axis field. Curves corresponding to successive temperatures are shifted by $\Delta L/L = 0.1\times 10^{-5}$ for visibility. All data were collected in up sweeps.}
	 \label{Fig:Data_Hiic}
\end{figure}

\begin{figure}[]
	\centering
	\includegraphics[width = 1\linewidth]{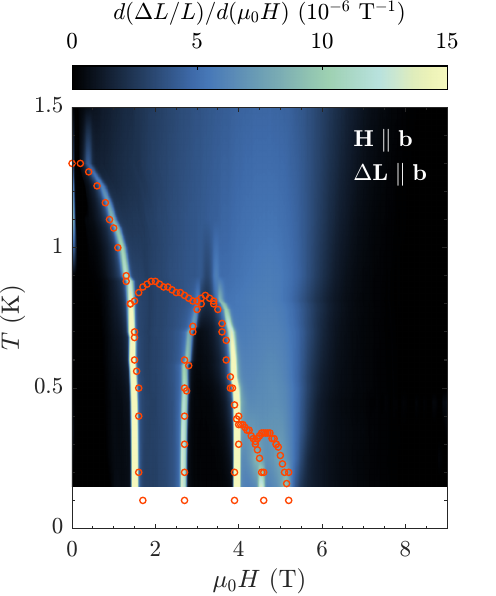}
	\caption{False-color plot of the field-derivative of the isothermal relative $b$-axis elongation, based on the actual data shown in Fig.~\ref{Fig:Data_Hiib}. Above 0.5~K, interpolation is performed in coordinates aligned with the local phase boundaries, allowing the sharp features to evolve continuously rather than fade between consecutive field scans. Symbols are phase boundaries deduced from specific heat measurements~\cite{Cs2CoBr4_PovarovPRR2020}.}
	 \label{Fig:PhaseDiagram_Hiib}
\end{figure}

\begin{figure}[]
	\centering
	\includegraphics[width = 1\linewidth]{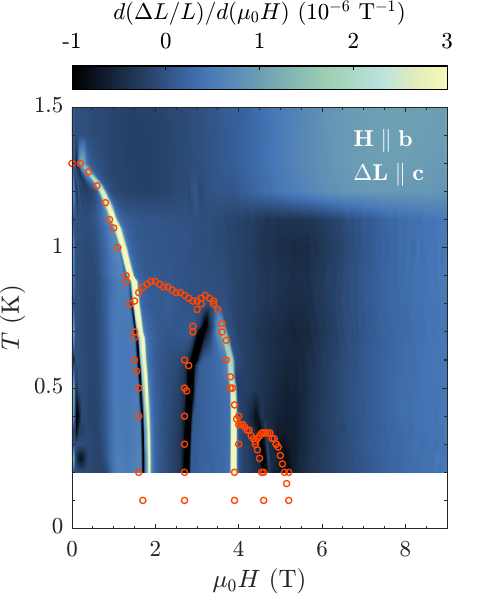}
	\caption{Same as Fig.~\ref{Fig:PhaseDiagram_Hiib} for elongation along the $c$-axis.}
	 \label{Fig:PhaseDiagram_Hiic}
\end{figure}

Dilatometry data are presented in Figs.~\ref{Fig:Data_Hiib} and \ref{Fig:Data_Hiic} as relative $b$- and $c$-axis isothermal elongation curves, respectively, plotted against the applied magnetic field. Distinct anomalies (small arrows) are observed at each of the phase boundaries in both geometries, at least at the lowest temperatures. Relative elongation along the $b$-axis is a monotonic function, spanning $\approx 2\times 10^{-5}$ over the experimental field range. The N\'eel, UUD and pseudospin-saturated phases are characterized by distinct plateaus. In the SDW, spin-flop and fan states the observed behavior is linear. In the studied range, relative elongation along the $c$-axis is over an order of magnitude smaller. Anomalies at phase boundaries remain clear, but the general monotonic increase of the kind seen along $b$ is absent. In both geometries the observed anomalies are particularly conspicuous in the field-derivatives of the elongation curves. The latter are presented in false-color plots in Figs.~\ref{Fig:PhaseDiagram_Hiib} and \ref{Fig:PhaseDiagram_Hiic}. The overlaid symbols are phase boundaries deduced from specific heat data~\cite{Cs2CoBr4_PovarovPRR2020}. We see that the entire phase diagram can be recovered from the dilatometry measurements alone.

\begin{figure}[]
	\centering
	\includegraphics[width = 1\linewidth]{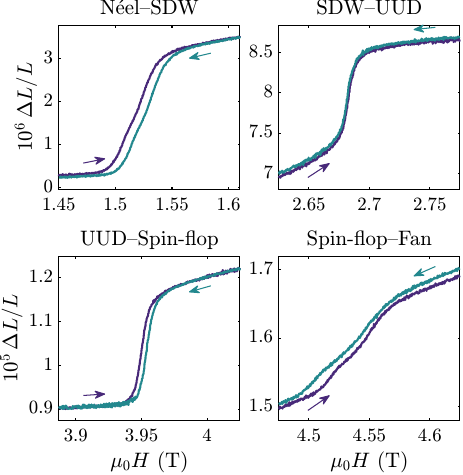}
	\caption{$b$-axis elongation measured in bidirectional field sweeps across phase transitions between the magnetically ordered phases at $T=0.15$~K. In all cases the sweep rate is $2\times 10^{-4}$~T/s.}
	 \label{Fig:dilhyst}
\end{figure}

Previous neutron diffraction experiments revealed the discontinuous nature of all but the highest-field transition \cite{Cs2CoBr4_FacherisPRB2024}, but did not provide sufficient accuracy to study hysteretic effects. Our dilatometry measurements fill this gap. Figure~\ref{Fig:dilhyst} zooms in on the measured $b$-axis elongation at the boundaries between the ordered phases. Clear hysteretic loops are observed for the N\'eel--SDW, UUD--spin-flop and spin-flop--fan transitions. The SDW--UUD incommensurate-to-commensurate transition is hysteresis-free.


\subsection{Ultrasound}

\begin{figure}[]
	\centering
	\includegraphics[width = 1\linewidth]{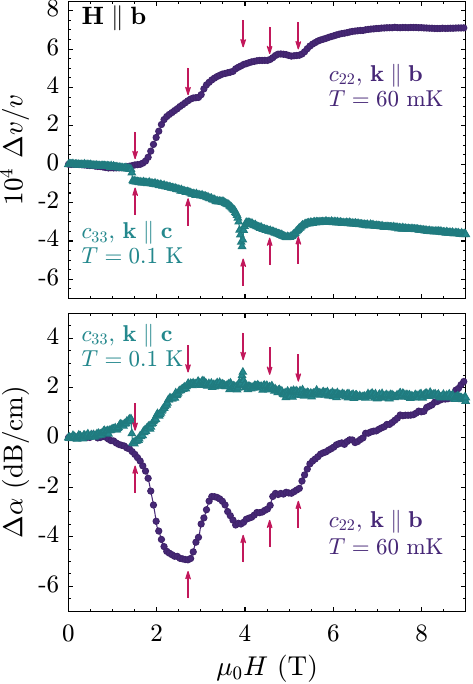}
	\caption{Field dependence of longitudinal ultrasound relative velocity changes (top) and attenuation (bottom) measured at base temperature for the $c_{22}$ and $c_{33}$ modes. Red arrows as in Fig.~\ref{Fig:Data_Hiic}.}
	\label{Fig:basetemp}
\end{figure}

The longitudinal sound velocities at room temperature were determined to be $v_{22}=2190(70)~\text{m}/\text{s}$ and $v_{33}=2110(20)~\text{m}/\text{s}$, parallel and perpendicular to the chain axis, respectively.
Figure~\ref{Fig:basetemp} shows the measured relative changes in sound velocity $\Delta v/v$ (top) and attenuation $\Delta \alpha$ (bottom) of both modes as a function of magnetic field applied along the $b$ axis. For both directions these quantities are of the order of $10^{-4}$--$10^{-3}$ and 5~dB/cm across the studied field range, respectively. 

Sound velocity anomalies are observed at most phase boundaries. The velocity of the chain-axis mode increases almost monotonically with increasing field. Each of the phase transitions is marked by a kink (change of slope) in the curve, the strongest one above the N\'eel phase. In contrast, that of the $c_{33}$ mode decreases. Here the upper boundary of the N\'eel phase shows a sharp jump, and that of the UUD a sharp spike. The saturation transition is marked by a kink, and only weak features can be guessed at the SDW--UUD and spin-flop--fan transitions. 

Ultrasound attenuation also shows roughly opposite behavior for the two propagation directions. For $c_{22}$ it drops sharply in the SDW state. The UUD plateau is marked by a broad peak. Beyond that, attenuation increases more or less continuously, to pseudospin saturation and beyond, marking each phase transition with a slope change. Attenuation for the $c_{33}$ mode increases through the SDW state, then decreases slightly to saturation and flattens out beyond that. The transition to the spin-flop phase is marked by a clear spike. The onsets of the fan and pseudospin-saturated states are practically unmarked.

\begin{figure*}[]
    \centering
    \includegraphics[width=\textwidth]{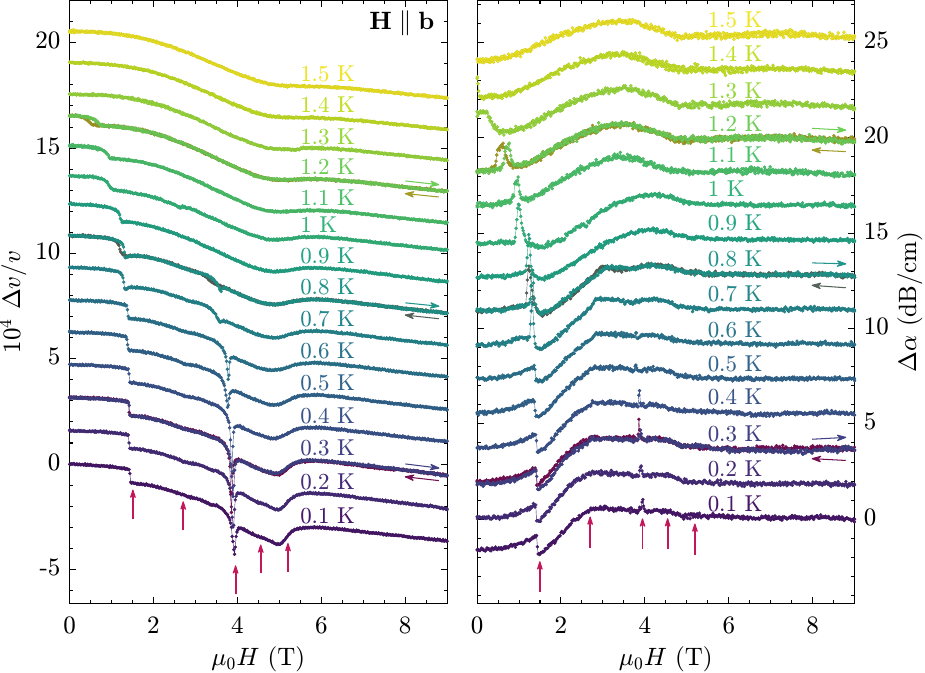}
    \caption{Measured relative sound velocity change $\Delta v/v$ (left) and change in ultrasound attenuation $\Delta \alpha$ (right) for the $c_{33}$ mode. The magnetic field is applied along the $b$-axis. Curves measured at different temperatures are offset for clarity. Arrows indicate the field sweep direction where both up and down sweeps are shown. Red vertical arrows indicate the low-temperature phase boundaries.}
    \label{fig:sound_stacked}
\end{figure*}

\begin{figure}[]
	\centering
	\includegraphics[width = 1\linewidth]{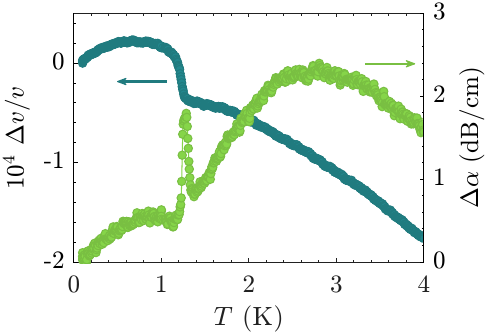}
	\caption{Temperature dependence of the relative change of the $c_{33}$ sound velocity (left axis) and of the corresponding attenuation (right axis) measured in zero field.}
	\label{Fig:Tscan_ultra}
\end{figure}

\begin{figure}[]
	\centering
	\includegraphics[width = 1\linewidth]{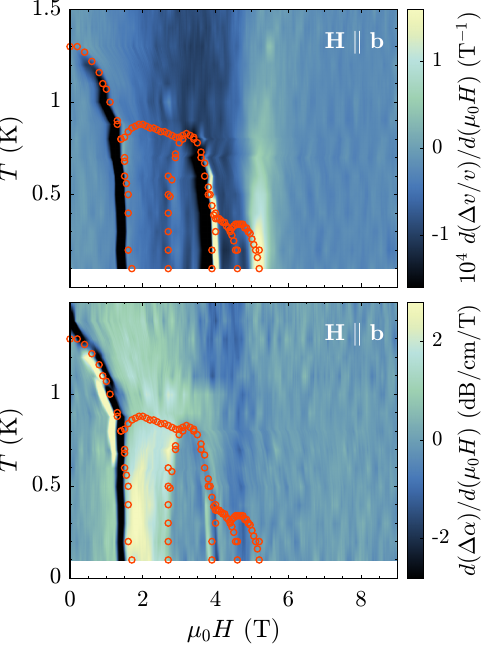}
	\caption{False-color plots of the field-derivatives of the measured relative $c_{33}$ sound velocity change (top) and attenuation (bottom), constructed from the field scans partially shown in Fig.~\ref{fig:sound_stacked}. As in Fig.~\ref{Fig:PhaseDiagram_Hiib}, above 0.5~K the interpolation is performed in coordinates aligned with the local phase boundaries. Symbols are as in Fig.~\ref{Fig:PhaseDiagram_Hiib}.}
	\label{Fig:soundcolor}
\end{figure}

The behavior of the $c_{33}$ mode was also studied as a function of temperature. Field scans at different temperatures are shown in 
Fig.~\ref{fig:sound_stacked}. The observed anomalies indeed follow the corresponding phase boundaries. The temperature dependencies measured in zero field are shown in Fig.~\ref{Fig:Tscan_ultra}. Magnetic ordering at $T_\text{N}=1.28(5)$~K is marked by a steep step in $\Delta v/v$ and a very sharp near-divergent spike in $\Delta\alpha$. The combined data sets for $c_{33}$ are presented in Fig.~\ref{Fig:soundcolor} as false-color plots of the field-derivatives of the sound velocity and attenuation, in analogy with the dilatometric maps of Figs.~\ref{Fig:PhaseDiagram_Hiib} and \ref{Fig:PhaseDiagram_Hiic}. Just as with the dilatometry data, the entire phase diagram can be read off the ultrasound measurements.

\begin{figure}[]
	\centering
	\includegraphics[width = 1\linewidth]{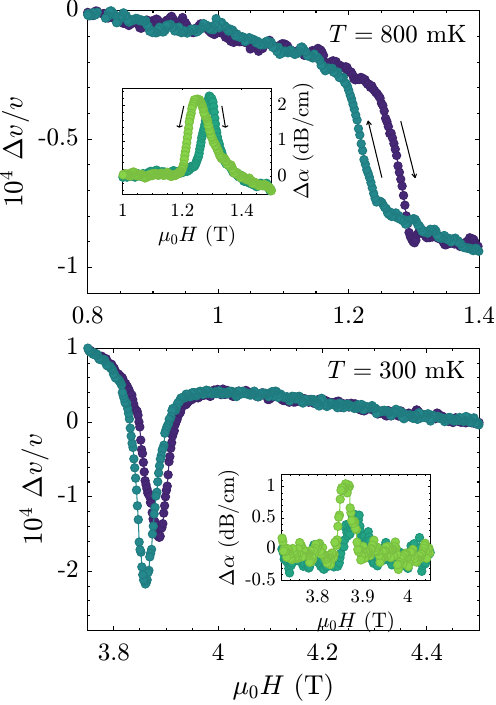}
	\caption{Relative change in the $c_{33}$ sound velocity (main panels) and attenuation (insets) measured in bidirectional field sweeps across the N\'eel--SDW (top) and UUD--spin-flop (bottom) phase transitions.}
	\label{Fig:hyster}
\end{figure}

Figure~\ref{Fig:hyster} shows the $c_{33}$ sound velocity (main panels) and attenuation (insets) measured in bidirectional sweeps across the N\'eel--SDW and UUD--spin-flop phase boundaries, where ultrasound anomalies are strongest. The data were taken with a deliberately slow field sweep rate of $10^{-4}$~T/s, to exclude any non-equilibrium effects. Consistent with the dilatometry results, hysteretic behavior is clearly visible. For the remaining phase transitions any hysteretic behavior is barely discernible or absent in our acoustic data. In particular, no history dependence is observed at the saturation transition.


\section{Discussion}
\label{sec:discussion}

\subsection{Order of the phase transitions and hysteresis}
\label{sec:hysteresis}
As already mentioned, the discontinuous nature of all but the saturation transition has previously been established in neutron diffraction studies, and is confirmed by the present measurements. Given the strong magnetoelastic coupling that we observe, the discontinuities are likely themselves due to interactions of the spin and lattice subsystems. Indeed, strain couples to the square of the order parameter, thus giving a negative contribution to the corresponding Landau free energy term \cite{BeanRodbell1962_PR126_104,LarkinPikin1969_JETP29_891}. If strong enough, it turns the initially continuous transition first order. Just one modern example of this scenario is
the study of discontinuous transitions in (Ca,Sr)Mn$_2$P$_2$ \cite{Pakhira2025_PRB112_104443_CaSrMn2P2_firstorder}. This said, most of the observed transitions would remain discontinuous even on a rigid lattice, as they are between order parameters of different symmetry. Most probably the lattice acts as a reinforcer of their discontinuous nature.

Discontinuous field-induced transitions in magnetic materials do not always involve hysteretic behavior. For example, the classic spin-flop transition in uniaxial antiferromagnets shows little or no hysteresis. It occurs through an intermediate state of coexisting domains, with only a very small metastability barrier \cite{KingPaquette1973_PRL30_662_MnF2_spinflop_domains,Bogdanov2007_PRB75_094425_spinflop_multidomain}. As mentioned above, in \CsCoBr only the SDW--UUD transition seems to be hysteresis-free, the other discontinuous transitions clearly show metastability.

The ultrasound data of Fig.~\ref{Fig:Tscan_ultra} show that, in contrast to the field-induced transitions, the temperature-driven ordering transition at the N\'eel point is a continuous one. As in classic examples of antiferromagnetic ordering \cite{LuthiMoranPollina1970_JPCS31_1741_sound_magnetic_transitions,Luthi2007}, the asymmetric spike ($\lambda$-anomaly) in attenuation is due to order parameter fluctuations and their critical slowing down at the transition point. Below the transition, fluctuations are rapidly suppressed as the order stiffens. Above $T_\text{N}$ one-dimensional correlations persist beyond the critical region, showing a broad peak around 2.5--3~K, a temperature scale set by the magnitude of $J_4$. The $\lambda$-anomaly at $T_\text{N}$, as well as the accompanying step in $\Delta v/v$, are far stronger than the barely resolvable features at $T_\text{N}$ in the sister compound \CsCuCl \cite{Cong2016_JAP120_142113_Cs2CuCl4_magnetoacoustic}, indicating a much stronger magnetoelastic coupling in the Co$^{2+}$ species.

\subsection{Exchange-striction model}

Exchange striction, the simple idea that for small deformations the exchange constants linearly depend on the lattice stress~\cite{Lines1979_PhysRep55_133_elastic_properties,Luthi2007}, appears to be the most likely origin of magnetoelastic coupling in \CsCoBr. At first glance, constructing a detailed microscopic model for our material presents a formidable challenge. The magnetic Hamiltonian features at least five relevant coupling constants, most of which involve multiple potential superexchange pathways. The difficulty can be overcome by leveraging the quasi-one-dimensional nature of the system, similarly to how it was previously done for DTN~\cite{Zapf2008_PRB77_020404_authorcopy}. Indeed, in \CsCoBr the $J_4$ bond is dominant and parallel to the crystallographic $b$ axis where the observed striction is largest. Moreover, the Co--Br--Br--Co superexchange (Fig.~\ref{Fig:Material_Structure}) that mediates it can be expected to be highly sensitive to $b$-axis pressure that will directly couple to the Br--Br distance and the crucial Co--Br--Br bond angle. Assuming this effect to be dominant, following the standard exchange-striction argument \cite{Lines1979_PhysRep55_133_elastic_properties,Anfuso2008_PRB77_235113} the $b$-axis elongation $\Delta L$ can be related to the exchange energy between nearest-neighbor pseudospins along the $b$ axis, $\mb{S}_1$ and $\mb{S}_2$. The delicate point is that exchange constants in the pseudospin frame are not isotropic. They are related to the actual Heisenberg constants through the Schrieffer--Wolff projection \cite{SchriefferWolff1966_PR149_491,Bravyi2011_AnnPhys326_2793}. Although in all magnetically ordered phases of \CsCoBr the ordered moments are exactly or closely aligned to the local ionic easy planes, the in-plane anisotropy still matters. In terms of pseudospins the exchange energy per bond is $\tilde{J}_4 \mathcal{C}= \tilde{J}_4 \left[(1+\delta)\aver{S_1^y S_2^y}+\aver{S_1^\perp S_2^\perp}\right]$, where $\tilde{J}_4=4J_4$ is the projected pseudospin coupling, $y$ is the chain axis, the perpendicular direction is in the local easy plane, and $\delta \approx 0.6$ \cite{Cs2CoBr4_HalatiPRR2026}. 

In the experiment elongation is measured at a small constant uniaxial pressure applied to the sample by the dilatometer plates. As worked out in detail in Ref.~\cite{Lines1979_PhysRep55_133_elastic_properties,Anfuso2008_PRB77_235113}, it can be directly expressed through the derivative of the exchange energies, which in our case gives:
\begin{equation}
        \Delta L/L =\frac{1}{V_0}\frac{\partial \tilde{J}_4}{\partial p_b}\mathcal{C},
    \label{eq:equilibrium}
\end{equation}
where $V_0$ is the volume per formula unit and $p_b$ is the uniaxial pressure along the $b$ axis. In brief, this is derived from the Maxwell relation for strain and magnetization $M$
\begin{equation}
\frac{1}{L}\left(\frac{\partial L}{\partial H}\right)_{p_b} = -\mo \left(\frac{\partial M}{\partial p_b}\right)_H,
\label{eq:Maxwell}
\end{equation}
which follows from differentiating the Gibbs potential density, whose differential is $\dd g= -\mo M \dd H + \Delta L/L \, \dd p_b$, with respect to $H$ and $p_b$ in either order and equating the results. Integrating Eq.~\eqref{eq:Maxwell} over field gives
\begin{equation}
\frac{\Delta L}{L} = \frac{1}{V}\frac{\partial F_\text{mag}}{\partial p_b}.
\end{equation}
The magnetic free energy $F_\text{mag}$ depends on pressure only via the exchange constants in the Hamiltonian, which immediately gives Eq.~\eqref{eq:equilibrium} for our $J_4$-only model.
Simplistic as the latter may appear, below we shall demonstrate that it reproduces the measured $b$-axis elongation remarkably well, at the same time providing new insight into the magnetic correlations.

\begin{figure}[]
	\centering
	\includegraphics[width = 1\linewidth]{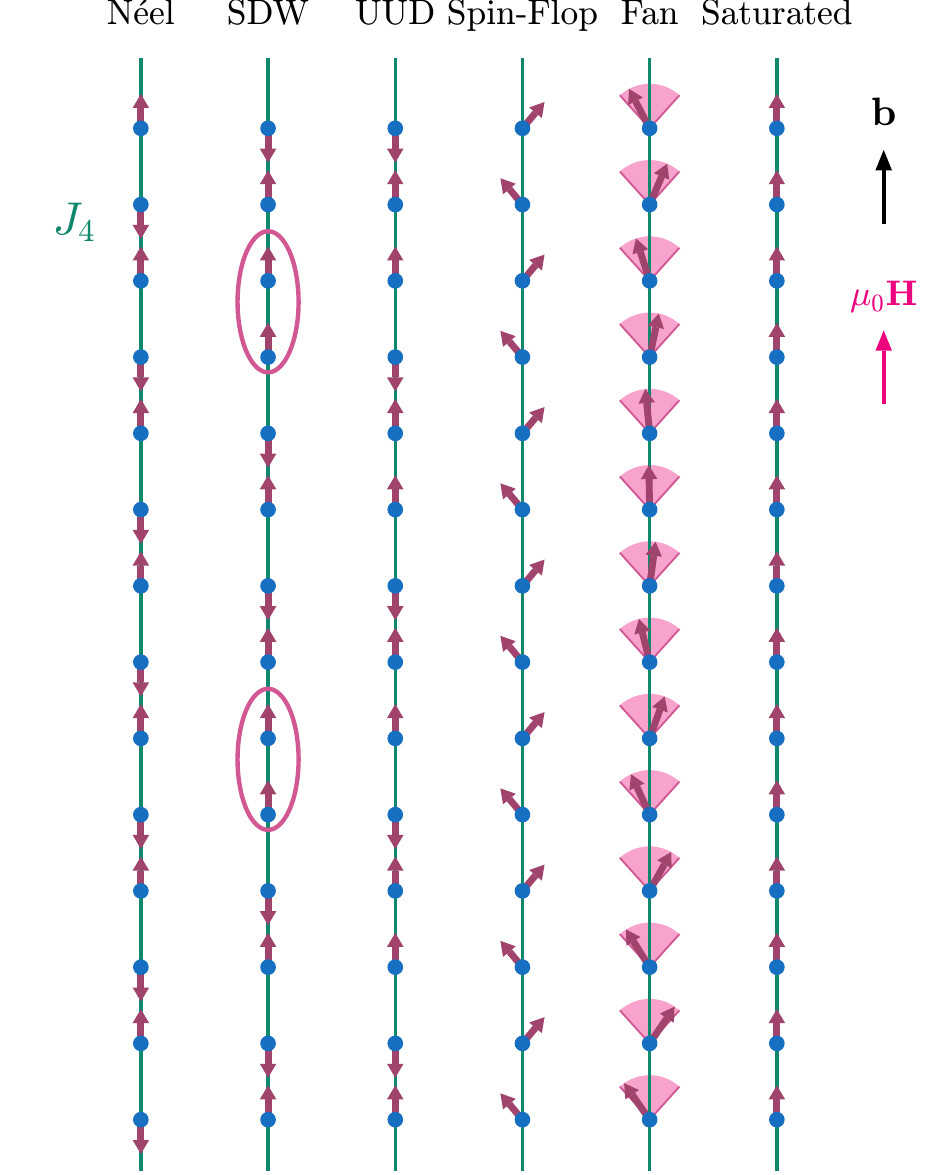}
	\caption{Schematic representation of the magnetic structures in \CsCoBr in magnetic fields applied along the $b$ axis. Only a single $b$-axis chain of cobalt ions, connected by the dominant bond $J_4$, is shown. The fan structure is simplified, omitting the tilt of the average spin direction with respect to the $b$ axis.}
	\label{Fig:PhaseSchematic}
\end{figure}

Using Eq.~\eqref{eq:equilibrium} to model the data is not as straightforward as it may seem. The simplest approach is to invoke the mean-field (MF) approximation, replacing $\aver{S_1^\alpha S_2^\alpha}\rightarrow \aver{S_1^\alpha}\cdot \aver{S_2^\alpha}$. This allows us to predict the elongation based on the magnetic structures previously determined by neutron diffraction \cite{Cs2CoBr4_FacherisPRL2022,Cs2CoBr4_FacherisPRB2024}. What makes the task considerably simpler is that according to those experiments the {\em magnitude} of the ordered magnetic moment remains almost field-independent (the deviation in the fan phase is primarily due to the field-induced Van Vleck admixture at high fields). In this rigid-pseudospin picture only the relative alignment changes across the phase diagram. This reduces the problem to computing the relative alignment of consecutive pseudospins in the $b$-chains. For each of the five ordered and the fully polarized phases illustrated in Fig.~\ref{Fig:PhaseSchematic} the exchange energy can be computed analytically.

The collinear N\'eel, UUD and SDW phases can be analyzed together. In the most complex of these, the SDW state, the magnitude of the propagation vector precisely follows pseudospin magnetization \cite{Cs2CoBr4_FacherisPRL2022}, similarly to the SDW phase in the prototype easy-axis antiferromagnetic spin chain material BaCo$_2$V$_2$O$_8$ \cite{Kimura2008_PRL101_207201_BaCo2V2O8_SDW,Canevet2013_PRB87_054408_BaCo2V2O8_diffraction}. Classically, it is described as an ordered array of domain walls (kinks) in the underlying N\'eel structure (see Fig.~\ref{Fig:PhaseSchematic}), each carrying a unit pseudospin magnetization, and is responsible for the so-called Devil's staircase in the long-range Ising model~\cite{BakBruinsma1982_PRL49_249_Ising_devils_staircase}. Quantum-mechanically the more appropriate language is that of incommensurate correlations in the Tomonaga--Luttinger spin liquid in the XXZ model \cite{Okunishi2007_PRB76_224411_XXZ_incommensurate}. Given that the emergent axial anisotropy due to crossed anisotropy planes in \CsCoBr is very strong, the kinks can be treated as localized on a single bond. In this picture the N\'eel state is a kink vacuum. In the UUD phase, the only step of the Devil's staircase visible, the kink concentration is $1/3$. A simple calculation shows that, averaged over the entire chain, the exchange correlator in these phases is given by:
\begin{equation}
	\mathcal{C}=S^2(1+\delta)(2\tilde{m}-1),
	\label{eq:mag_sdw}
\end{equation}
where $\tilde{m}=\aver{S^y}/S$ is the $b$-axis pseudospin magnetization. From Eq.~\eqref{eq:equilibrium} it then immediately follows that in these phases, to within a constant offset, $\Delta L/L$ can be directly read off the pseudospin magnetization curve. For BaCo$_2$V$_2$O$_8$, dilatometry~\cite{Niesen2013_PRB87_224413} and ultrasound~\cite{Yamaguchi2011_JPSJ80_033701_BaCo2V2O8_ultrasound} measurements have indeed revealed clear anomalies at the transition from the N\'eel to the field-induced SDW phase. What distinguishes \CsCoBr is that at higher fields it goes through non-collinear phases that are not captured by the kink picture.

In the spin-flop phase the math is even simpler: computing the spin tilt angle from $\tilde{m}$ we immediately get:
\begin{equation}
	\mathcal{C}=S^2\left[(2+\delta)\tilde{m}^2-1\right].
	\label{eq:mag_sf}
\end{equation}

The fan structure is the most complex of all, with spins oscillating with an amplitude $\phi\sim 35^\circ $ in a plane, around an axis tilted by some angle $\psi\sim 28^\circ $ with respect to the $b$ axis. These two values were determined at $\mo H=4.8$~T, and their field dependence remains unexplored. The best we can do is compute the exchange correlator for a simplified structure, assuming $\psi=0$. The magnetic propagation vector $(0,0.47,0)$ is very close to the antiferromagnetic point, so {\em locally} the structure resembles the spin-flop state, where 
\begin{equation*}
	\mathcal{C}=S^2\left[(2+\delta)\tilde{m}_\text{local}^2-1\right].
\end{equation*}
In the fan structure the local magnetization $\tilde{m}_\text{local}$ is spatially modulated over its long period. Averaging the $\cos$ function over this harmonic modulation gives $\tilde{m}=J_0(\phi)$ (zeroth-order Bessel function) and 
\begin{equation*}
	\mathcal{C} =S^2\left[\frac{(2+\delta)}{2}\bigl(1+J_0(2\phi)\bigr)-1\right].
\end{equation*}

For $\tilde{m}\rightarrow 1$, near saturation, an expansion gives
\begin{equation}
	\mathcal{C}=\frac{S^2}{2}\left[(2+\delta)(3\tilde{m}^2-2\tilde{m})+\delta\right].
	\label{eq:mag_fan}
\end{equation}

Finally, in the pseudospin saturation phase $\aver{\mb{S}}\parallel\mb{b}$, so that
\begin{equation}
	\mathcal{C}=(1+\delta)\,S^2\tilde{m}^2.
	\label{eq:mag_sat}
\end{equation}

\begin{figure*}[]
	\centering
	\includegraphics[width = \textwidth]{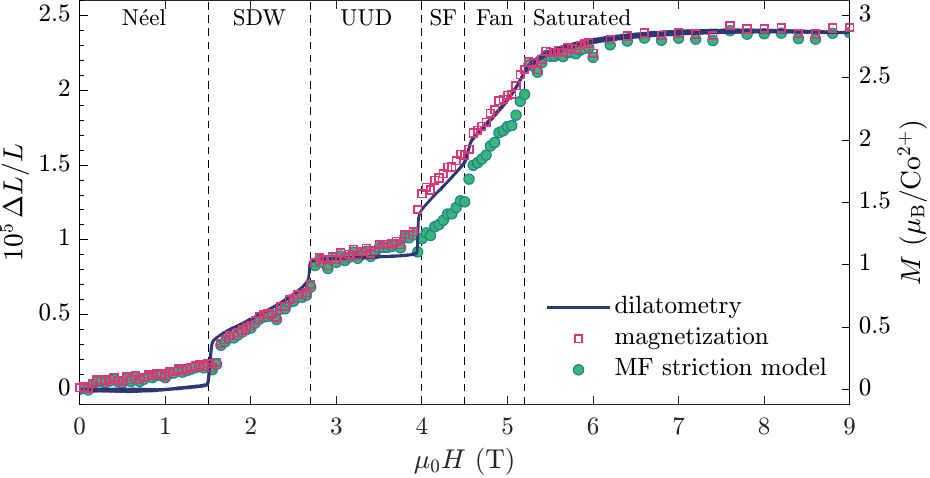}
	\caption{Field dependence of the relative $b$-axis elongation of \CsCoBr. Solid line: experimental data measured at $T=0.15$~K. Circles: mean-field exchange-striction model, as described in the text. Squares: magnetization measured at $T=0.1$~K with Van Vleck contribution subtracted, from Ref.~\cite{Cs2CoBr4_PovarovPRR2020}.}
	\label{Fig:Model}
\end{figure*}

Using the pseudospin magnetization data of Ref.~\cite{Cs2CoBr4_PovarovPRR2020} we can thus predict the elongation with only two unknown parameters: the overall offset and the scale factor. Computing these to match the elongation measured in zero field and at $\mo H=9$~T (deep in the pseudospin saturation state) we get the result plotted as filled circles in Fig.~\ref{Fig:Model}. For comparison, the experimental data are plotted as a solid line. Qualitative agreement is observed across the entire range and is particularly good in the N\'eel, SDW and UUD phases. At higher fields the model deviates from the data. It fails to follow the observed elongation steps, even as it gets the slope more or less right. Recall that in the three lower-field phases, where the model fits the data well, its prediction is simply proportional to magnetization. The central and rather striking result of this study is that elongation closely follows the magnetization (magenta squares in Fig.~\ref{Fig:Model}) through the {\em entire phase diagram}, even when it starts deviating from the simple-minded MF model.

To understand why elongation follows magnetization rather than its square at high field as well, one has to abandon the MF picture entirely. A rather general alternative argument can be made for the vicinity of saturation, where the paramagnetic state connects to the two transverse-ordered phases.
We shall postulate that in this regime, at a given temperature, the {\em leading-order} contribution to the free energy $F_\text{mag}$ of the system can be written as a function of $H_\text{sat}-H$ alone.
The saturation field $H_\text{sat}$ is in turn defined by the energy scale, in our case by $J_4$. Any shift of $J_4$ due to uniaxial pressure $p_b$ applied to the sample translates to a linear shift of $H_\text{sat}$, so that it has the same effect on quasiparticles as a field shift. In other words,
\begin{equation}
	\left(\frac{\partial M}{\partial H}\right)_{p_b}\propto \left(\frac{\partial M}{\partial p_b}\right)_H.
\end{equation}
This, combined with Eq.~\eqref{eq:Maxwell}, yields
\begin{equation}
	\left(\frac{\partial L}{\partial H}\right)_{p_b} \propto \left(\frac{\partial M}{\partial H}\right)_{p_b}.
\end{equation}
A dilatometer operates at a constant small pressure and therefore directly measures the quantity on the left-hand side. Integrating over field leads to
\begin{equation}
	L(H_\text{sat})-L(H)\propto \tilde{m}(H_\text{sat})-\tilde{m}(H).\label{eq:elongation}
\end{equation}

The assumption regarding free energy applies, for instance, in the axially symmetric case. Here magnetic order is described as a ``condensate'' of quasiparticles, i.e., magnons or spin flips \cite{BatyevBraginskii1984_JETP60_781_bose_gas_analogy}. The magnon density is given by $1-\tilde{m}$, while the magnetic order parameter (spontaneous transverse magnetization) is a measure of magnon coherence in the condensate. The magnon chemical potential is given by $\mu=g\mo\mub (H_\text{sat}-H)$. The physical meaning of Eq.~\eqref{eq:elongation} is that each condensing magnon (spin flip) causes a fixed local exchange bond-energy change, and therefore produces a fixed lattice displacement. Lattice contraction simply counts the number of spin flips, so that elongation scales with magnetization. Similar behavior is observed in the prototypical magnon Bose--Einstein condensate (BEC) material DTN \cite{Zapf2008_PRB77_020404_authorcopy} near its saturation transition at $H_{c2}$. Unlike in \CsCoBr though, in DTN elongation no longer tracks magnetization deeper into the ordered phase. There it becomes very nonlinear, eventually reversing the sign of $\dd L/\dd H$.  This is undoubtedly due to the fact that in DTN the ordered state does not extend all the way to zero field but is terminated at $H_{c1}$, below which the system is again paramagnetic. On this end, the condensing magnons are single-ion $\ket{S_z=0}\rightarrow \ket{S_z= 1}$ transitions, rather than spin flips. They play the opposite role in the sense that each magnon {\em increases} magnetization. As a result, in the vicinity of $H_{c1}$, DTN $\Delta L/L$ is again proportional to magnetization, but with a {\em negative} prefactor.

Our material \CsCoBr is not axially symmetric. Nevertheless, the magnon-counting picture may still work, at least approximately. The almost-orthogonal easy planes in adjacent spin chains may be expected to at least partially relieve off-axial anisotropy in the long-wavelength limit. This is supported by the fact that both transverse-ordered phases, unlike the N\'eel, UUD and saturated states, show gapless specific heat \cite{Cs2CoBr4_PovarovPRR2020}. In any case, the experimental fact remains: in \CsCoBr elongation is proportional to magnetization across the entire field range.

At high field the physics is similar to that in DTN near saturation, and can be discussed in terms of condensing magnons or (pseudo)spin-flips. At low fields, in the collinear phases, the behavior is not due to single-ion transitions as in DTN, but to topological kinks endemic to the quasi-one-dimensional character of \CsCoBr.

\subsection{Quantifying the magnetoelastic coupling}
The most important quantitative information that can be extracted from our measurements is the actual strength of magnetoelastic coupling for $J_4$. Between the N\'eel and pseudospin saturation phases, from the low-temperature data shown in Fig.~\ref{Fig:Data_Hiib}, $\Delta L/L$ changes by $2.40(2)\times 10^{-5}$. In the same range the correlator flips sign: $\mathcal{C}=\mp \, (1+\delta)S^2$. For $J_4$ in \CsCoBr, the relevant pseudospin parameter is $(1+\delta)\tilde{J}_4\approx 0.8~\text{meV}$ \cite{Cs2CoBr4_HalatiPRR2026}. Combining these values in Eq.~\eqref{eq:equilibrium} gives
\begin{align}
	\frac{\partial \ln \tilde{J}_4}{\partial p_b}&=\frac{2V_0}{(1+\delta)\tilde{J}_4}\left[\frac{\Delta L}{L}\bigg|_{H>H_\text{sat}}-\frac{\Delta L}{L}\bigg|_{H=0}\right]\nonumber\\
	&=0.10(1)~\text{GPa}^{-1}.
	\label{eq:coupling}
\end{align}
This value is to be compared against the established empirical Bloch's ``10/3 law'' for the volume dependence of superexchange \cite{Bloch1966_JPCS27_881_superexchange_pressure}. If we assume that $J_4$ is much less sensitive to $p_c$ and $p_a$, as supported by the discussion of $c$-axis data below, using the values for the bulk modulus $K$ reported for isostructural Cs$_2$CuCl$_4$ and Cs$_2$CoCl$_4$ \cite{Xu2000_JSSC153_212_Cs2CuCl4_Cs2CoCl4_pressure_XRD,Nataf2017_PRB95_014110_Cs2CoCl4_pressure_structure_Raman}, we get
\begin{equation}
	\frac{\partial \ln \tilde{J}_4}{\partial \ln V}=-K \frac{\partial \ln \tilde{J}_4}{\partial p_b}=-1.5~\text{to}~-2.1,
	\label{eq:couplingV}
\end{equation}
about half of the Bloch value. A perfect agreement is not to be expected: the latter was established for superexchange via a single anion, while in \CsCoBr the main interaction runs over a more complex two-anion bridge. A more appropriate comparison can be made with other species of the same structural family, where similar superexchange routes are involved. Indeed, for \CsCuCl hydrostatic pressure measurements give ${\partial \ln J}/{\partial \ln V}\sim -1.2$ \cite{Thallapaka2019_PSSb256_1900044_Cs2CuClBr_pressure}, even smaller than our estimate for \CsCoBr.

\subsection{Chain-axis ultrasound velocity}
\label{sec:ultrasound}
While the sample elongation in the $J_4$-only exchange striction model is more or less a direct measure of the exchange correlator $\mathcal{C}$, the relative change in longitudinal sound velocity is related to stiffness and has two contributions \cite{TachikiMaekawa1974_PTP51_1_sound_field_theory,Lines1979_PhysRep55_133_elastic_properties}. One term is due to the second derivative of the exchange constant
\begin{equation}
\frac{\Delta v^{(2)}}{v} = \frac{1}{2c_{22}V_0}\frac{\partial^2 \tilde{J}_4}{\partial \epsilon_{22}^2}\,\mathcal{C}
 = \frac{1}{2c_{22}}\,\frac{\partial^2 \tilde{J}_4/\partial \epsilon_{22}^2}{\partial \tilde{J}_4/\partial p_b}\;\frac{\Delta L}{L},
 \label{eq:v2}
\end{equation}
where $c_{22}=\rho\,v_{22}^2=19.3$~GPa is the corresponding component of the elastic tensor. Thus, the second-derivative term simply echoes the elongation $\Delta L/L$ in this model.
The ``first-derivative'' term is due to the ``exchange susceptibility'', i.e., the second $J$-derivative of the magnetic free energy $F_\text{mag}$, $\chi_{\tilde{J}_4}=-\frac{\partial^2 F_\text{mag}}{\partial \tilde{J}_4^2}$:
\begin{equation}
\frac{\Delta v^{(1)}}{v} = -\frac{1}{2c_{22}V_0}\left(\frac{\partial \tilde{J}_4}{\partial \epsilon_{22}}\right)^2 \chi_{\tilde{J}_4}.
\end{equation}

At low temperatures, if $J_4$ is assumed to be the only magnetism-relevant energy scale, the magnetic energy must have the form $F_\text{mag}=\tilde{J}_4 f(g\mub\mo H/\tilde{J}_4)$, where $f$ is some scaling function. Differentiating twice with respect to $\tilde{J}_4$ gives:
\begin{equation}
\chi_{\tilde{J}_4} = (\mo H/\tilde{J}_4)^2\chi,
\end{equation}
where $\chi$ is the magnetic susceptibility per formula unit. Thereby, for $T\rightarrow 0$ (cf. Refs.~\cite{TachikiMaekawa1974_PTP51_1_sound_field_theory,Chiatti2008_PRB78_094406_DTN_magnetoacoustics,Zherlitsyn2014_LTP40_123_spinlattice_AFM_review}),
\begin{equation}
\frac{\Delta v^{(1)}}{v} = -\left(\frac{\partial \ln \tilde{J}_4}{\partial \epsilon_{22}}\right)^2 \frac{(\mo H)^2}{2c_{22}V_0}\,\chi.
\end{equation}

We can estimate $\partial \ln \tilde{J}_4/\partial \epsilon_{22}$ using our result in Eq.~\eqref{eq:coupling}:
\begin{equation}
\frac{\partial \ln \tilde{J}_4}{\partial \epsilon_{22}}
\approx -c_{22}\,\frac{\partial \ln \tilde{J}_4}{\partial p_b}\approx -1.9.
\end{equation}
Then, noting that $\chi$ never exceeds $1~\mub/\text{T}$ except for the discontinuities at the phase transitions, we can put a conservative upper bound on the first-derivative contribution in the $c_{22}$ mode:
\begin{equation}
\left|\frac{\Delta v^{(1)}_{22}}{v_{22}}\right| \lesssim 10^{-4},
\end{equation}
which is much smaller than the observed overall magnitude of the effect. To a large degree, the observed behavior is then to be attributed to the second-derivative term. Indeed, the measured $\Delta v_{22}/v_{22}$ and $\Delta L/L$ are both more or less flat in the N\'eel and saturated phases, and grow monotonically in-between. From zero field to saturation $v_{22}$ increases by a relative $7\times 10^{-4}$. The magnitude of the effect is the same as in DTN \cite{Chiatti2008_PRB78_094406_DTN_magnetoacoustics}, where $\Delta L/L$ along the chain bond is an order of magnitude larger \cite{Zapf2008_PRB77_020404_authorcopy}. The implication is that in \CsCoBr the curvature of $\tilde{J}_4$ is much larger than its slope. From Eq.~\eqref{eq:v2} we can then estimate
\begin{equation}
\frac{1}{\tilde{J}_4}\,\frac{\partial^2 \tilde{J}_4}{\partial \epsilon_{22}^2}=2c_{22}\,\frac{\partial \ln \tilde{J}_4}{\partial p_b}\,\frac{\Delta v_{22}/v_{22}}{\Delta L/L}\approx 10^2,
\label{eq:coupling2}
\end{equation}
in the ballpark of what an $r^{-10}$ Bloch-type bond would have \cite{Bloch1966_JPCS27_881_superexchange_pressure}.

The finer details of the measured field-dependence of $\Delta v_{22}/v_{22}$ are not reproduced by the second-derivative term Eq.~\eqref{eq:v2}. In particular, the measured velocity grows across the plateau and is largely flat in the transverse-ordered phases, while elongation does exactly the opposite. This shows the limitations of the simplistic single-$J_4$ approach to describing $b$-axis sound propagation in this material. Other exchange constants and the strain coupling of the anisotropy term may also play a role.

\subsection{$c$-axis elongation and ultrasound velocity}
\label{sec:caxis}

Dilatometry transverse to the $b$ axis could, in principle, bring insight into the other, weaker bonds in the system. Indeed, to a first approximation, $J_4$ is expected to be decoupled from $p_c$. By symmetry, a $c$-axis strain can only rotate Br$_4$ coordination tetrahedra of Co$^{2+}$ around the $b$ axis. Yet, these tetrahedra are known to be far more rigid than any other bonds in this family of materials \cite{Xu2000_JSSC153_212_Cs2CuCl4_Cs2CoCl4_pressure_XRD,Nataf2017_PRB95_014110_Cs2CoCl4_pressure_structure_Raman}. This makes the shape of the Co--Br--Br--Co superexchange pathway largely insensitive to transverse distortions. This is fully consistent with the measured $\Delta L/L$ for the $c$-axis: Fig.~\ref{Fig:Data_Hiic} shows no trace of the $J_4$ response seen along $b$, even at a 1~\% level. Any $c$-axis elongation is thus to be attributed to either the two next-strongest bonds, $J_1$ and $J_5$, or the magnetoelastic effects involving single-ion anisotropy \cite{Lines1979_PhysRep55_133_elastic_properties}, the latter, we recall, being the largest magnetic energy scale. Untangling these individual contributions does not appear feasible.
Overall, the overwhelming dominance of magnetic field induced elongation along the chain axis over that in transverse directions is not uncommon and is also observed in DTN \cite{Zapf2007_JAP101_09E106_BECstriction,Zapf2008_PRB77_020404_authorcopy}.

Turning to ultrasound, the one experimental result that stands out is the sheer magnitude of the relative $c$-axis velocity change. In this geometry the relative elongation across the magnetic phase diagram is about 20 times smaller than it is for $b$, but the changes in sound velocities are comparable in magnitude. The difference of behavior along the two axes is of course not surprising, as entirely different bonds are involved. If we attribute the large $\Delta v_{33}/v_{33}$ variation entirely to the second-derivative term, we would need to assume curvatures an order of magnitude larger than in Bloch's law for the relevant bonds, be they $J_1$ or $J_5$. Indeed, the order of magnitude of $\partial ^2 J/ \partial \epsilon ^2$ needed to produce the observed $\Delta v_{33}/v_{33}$ via the second-derivative term would be similar to that for the $b$-axis. At the same time, the exchange constants themselves are half an order of magnitude smaller than $J_4$. The observed effect is therefore likely due to the first-derivative term, which is consistent with its negative sign, i.e., with the observed field-induced softening of the lattice \cite{TachikiMaekawa1974_PTP51_1_sound_field_theory,Lines1979_PhysRep55_133_elastic_properties}. As for the $b$-axis, anisotropy striction may also be expected to be relevant. The field-induced admixing of the higher-energy Co$^{2+}$ Kramers doublet must also play a role, as it alone can account for the residual velocity slope in the pseudospin-saturated phase.

\subsection{Ultrasound attenuation}
Ultrasound attenuation is considerably harder to model than velocity, since it is a kinetic rather than a thermodynamic quantity. It may be expected to represent sound dissipation via magnon creation, and should be lowest in the gapped phases. The observed behavior deviates starkly from this naive picture. Attenuation of the $c_{22}$ mode is actually smaller in the gapless phases than in the N\'eel state. The gapped plateau shows an attenuation maximum. In the pseudospin-saturated state it is the largest and continues to increase with field, even as the spin gap increases. For $c_{33}$ the only large-scale feature is a stepwise increase of attenuation across the incommensurate phase. Unfortunately, the only conclusion to draw from these data is that all field-induced transitions are characterized by attenuation anomalies.

\section{Conclusion}
The entire cascade of field-induced magnetic transitions in \CsCoBr is witnessed by the underlying crystal lattice. Its response is manifested in striction and sound-propagation anomalies, showing discontinuities at all boundaries between ordered phases and hysteretic behavior at most of them. 

Thanks to the quasi-one-dimensional nature of the material, the observed phenomena can be largely attributed to exchange striction in a single exchange bond $J_4$ that forms the spin chains. Remarkably, sample elongation along that direction closely tracks magnetization across the entire phase diagram. In other words, it tracks the density of quasiparticles ``condensed'' into the ground state by the applied field. In the Ising-like collinear phases at low fields these quasiparticles are kinks (antiphase domain walls) in the underlying N\'eel structure, each carrying a unit magnetization and contributing a fixed amount to lattice dilation. Their contribution to striction is accounted for by a simple mean-field treatment. In contrast, the appropriate language for the transverse-ordered magnetic-BEC phases at high fields is that of magnons. Each such quasiparticle reduces the magnetization by unity and produces a fixed lattice contraction. 

The strength of magnetoelastic coupling for the $b$-direction, where it can be quantitatively analyzed, is somewhat smaller than given by Bloch's law. We attribute this to the geometry of the superexchange pathways that the chain-axis coupling involves. Remarkably, magnetic striction in the transverse direction is one and a half orders of magnitude smaller than along $b$, even as field-induced changes to sound velocity are comparable in magnitude.

\begin{acknowledgments}
Some single-crystal samples used in this work were previously prepared by Dr. Leonardo Facheris and Dr. Zewu Yan (ETHZ). Work at ETHZ was partially supported by the Swiss National Science Foundation, MINT grant no.~200021\_212753. This work was supported by the Deutsche Forschungsgemeinschaft through the W\"{u}rzburg-Dresden Cluster of Excellence on Complexity and Topology in Quantum Matter -- ct.qmat (EXC 2147, project No.~390858490), as well as by the HLD at HZDR, member of the European Magnetic Field Laboratory (EMFL).

\end{acknowledgments}

\bibliographystyle{apsrev4-2}
\bibliography{bibliography}
	
\end{document}